\input ./style/arxiv-general.cfg
\documentclass[aoas,MSNbibl,nameyear,seceqn,dvips]{arximspdf}
\makeatletter
   \@ifpackageloaded{graphicx}{}{\usepackage{graphicx}}
\makeatother
\usepackage{dcolumn,multirow}

\doi{10.1214/15-AOAS824}
\volume{9}
\issue{2}
\pubyear{2015}
\firstpage{547}
\lastpage{571}
\docsubty{FLA}

\makeatletter
\newcolumntype{d}[1]{D{.}{.}{#1}}

\newcommand{\eqref}[1]{(\ref{#1})}
\makeatother

\begin{document}
\begin{frontmatter}

\title{Modeling sea-level change using errors-in-variables integrated Gaussian processes\thanksref{T1}}
\runtitle{Modeling sea level change}
\thankstext{T1}{Supported in part by the Structured Ph.D. in Simulation
Science which is funded by the Programme
for Research in Third Level Institutions (PRTLI) Cycle 5 and co-funded
by the
European Regional Development Fund, and the Science Foundation Ireland
Research Frontiers Programme (2007/RFP/MATF281)
and also supported by the National Oceanic and Atmospheric
Administration award NA11OAR4310101
and NSF award EAR1402017 and OCE1458904. This paper is a contribution to PALSEA2.}

\begin{aug}
\author[A]{\fnms{Niamh} \snm{Cahill}\corref{}\thanksref{T1}\ead[label=e1]{niamh.cahill.2@ucdconnect.ie}},
\author[B]{\fnms{Andrew C.} \snm{Kemp}\thanksref{T2}\ead[label=e4]{andrew.kemp@tufts.edu}},
\author[C]{\fnms{Benjamin~P.}~\snm{Horton}\thanksref{T3,T4}\ead[label=e3]{bphorton@marine.rutgers.edu}}\\
\and
\author[A]{\fnms{Andrew C.} \snm{Parnell}\thanksref{T1}\ead[label=e2]{andrew.parnell@ucd.ie}}
\runauthor{Cahill, Kemp, Horton and Parnell}
\affiliation{University College Dublin\thanksmark{T1},
Tufts University\thanksmark{T2},
Rutgers University\thanksmark{T3}  and
Nanyang Technological University\thanksmark{T4}}
\address[A]{N. Cahill\\
A. C. Parnell\\
School of Mathematical Sciences\\
Complex and Adaptive Systems Laboratory\\
Earth Institute\\
University Collge Dublin\\
Belfield\\
Dublin 4\\
Ireland\\
\printead{e1}\\
\phantom{E-mail:\ }\printead*{e2}}
\address[B]{A. C. Kemp\\
Department of Earth and Ocean Sciences\\
Tufts University\\
2 North Hill Road\\
Medford, Massachusetts 02155\\
USA\\
\printead{e4}}
\address[C]{B.~P. Horton\\
Department of Marine and Coastal Sciences\\
\quad and Institute of Earth, Ocean\\
\quad and Atmospheric Sciences\\
Rutgers University\\
71 Dudley Road\\
New Brunswick, New Jersey 08901\\
USA\\
and\\
The Earth Observatory of Singapore\\
\quad and Asian School of the Environment\\
Nanyang Technological University\\
50 Nanyang Ave\\
Singapore 639798\\
\printead{e3}}
\end{aug}

%
\received{\smonth{7} \syear{2014}}
%
\revised{\smonth{3} \syear{2015}}

%
\begin{abstract}
We perform Bayesian inference on historical and late Holocene (last
2000 years) rates of sea-level change. The input data to our model are
tide-gauge measurements and proxy reconstructions from cores of coastal
sediment. These data are complicated by multiple sources of
uncertainty, some of which arise as part of the data collection
exercise. Notably, the proxy reconstructions include temporal
uncertainty from dating of the sediment core using techniques such as
radiocarbon. The model we propose places a Gaussian process prior on
the rate of sea-level change, which is then integrated and set in an
errors-in-variables framework to take account of age uncertainty. The
resulting model captures the continuous and dynamic evolution of
sea-level change with full consideration of all sources of uncertainty.
We demonstrate the performance of our model using two real (and
previously published) example data sets. The global tide-gauge data set
indicates that sea-level rise increased from a rate with a posterior
mean of 1.13 mm$/$yr in 1880~AD (0.89 to 1.28 mm$/$yr 95\% credible interval
for the posterior mean) to a posterior mean rate of 1.92 mm$/$yr in
2009~AD (1.84 to 2.03 mm$/$yr 95\% credible interval for the posterior
mean). The proxy reconstruction from North Carolina (USA) after
correction for land-level change shows the 2000~AD rate of rise to have
a posterior mean of 2.44 mm$/$yr (1.91 to 3.01 mm$/$yr 95\% credible
interval). This is unprecedented in at least the last 2000 years.
\end{abstract}

%
\end{frontmatter}

\section{Introduction}\label{sec1}
Sea-level rise poses a hazard to the intense concentrations of
population and infrastructure that are increasingly located at the
coast [\citet{Nicholls2010}]. Effective mitigation and management of
this hazard is reliant upon accurate estimation of historic, current,
and future rates of sea-level rise. Data for estimating such rates come
from instrumental measurements (tide gauges and satellites) and proxy
reconstructions (derived from a wide variety of palaeoenvironmental
data including stratigraphical,
biological, geochemical, and archaeological data).
Instrumental data are more precise, but span the relatively short
historic time period. Proxy reconstructions are less precise, but cover
a much longer time interval. We use examples of both types of data to
estimate rates of sea-level change with thorough quantification of
uncertainty.

The instrumental data we use provides a historic time series of fixed
and known ages with estimated sea levels and associated measurement
errors. Although there are now $\sim$2000 operational tide gauges
worldwide [\citet{JJ06,Player03}] that are located along coastlines and
islands, most were installed since the 1950s. Therefore, global
compilations rely on fewer gauges further back in time. The most widely
used global tide-gauge compilation spans the period since 1880~AD [\citet
{CandW2011}]. Since late 1992~AD, satellite altimetry measurements have
further provided a global record of sea-level change [\citet
{Nerem2010,Cazenave10}]. \citet{CandW2011} demonstrate that there is
good agreement (within uncertainty bounds) between their global mean
sea-level (GMSL) record based on tide gauges and satellite altimetry
measurements over the period from 1993~AD to 2009~AD. Thus, we use only
the tide-gauge data as our instrumental record.

Proxy data provide sea-level reconstructions spanning hundreds to
millions of years. Here we use late Holocene (last 2000 years) data to
place modern rates of sea-level change in an appropriate context and
characterize the  relationship between climate and sea
level. In our case study, we use proxy data that were preprocessed from
their raw form (counts of  species preserved within cores of coastal
sediment) into estimates of sea level. We do not explore the
preprocessing in this paper; see \citet{Birks95,Horton06,Juggins2012}
for a discussion of how this was done. The resulting processed data are
comprised of sea-level estimates that are irregularly spaced in time
and have uncertain ages in addition to sea-level uncertainties.

Instrumental and proxy reconstructions both estimate relative sea level
(RSL), which is the product of simultaneous land- and ocean-level
changes. In the absence of tectonics, land-level changes primarily
arise from the ongoing, slow rebound of the solid Earth to deglaciation
[\citet{Peltier04}], which is called glacio-isostatic adjustment (GIA).
Regions that were under the thickest ice at the last glacial maximum
(between 26,000 and 19,000 years ago) are experiencing uplift (RSL
fall), while areas that were peripheral to the ice sheet are
experiencing subsidence (RSL rise). To compare sea-level measurements
or reconstructions from different locations and to isolate the
climate-related component of sea-level change, it is necessary to
estimate and remove the contribution from GIA [\citet{Engelhart09}]. The
global tide-gauge data set that we use in this paper was already corrected
for GIA [\citet{CandW2011}], but we must correct the proxy reconstruction.
Since GIA is a rate (usually expressed in mm$/$yr), it affects older
sediments more than younger sediments. This has repercussions for our
model, because it introduces correlation between the individual age and
sea-level reconstructions. We defer full discussion of this to
Section~\ref{sec4}.

To accurately estimate the evolution of rates of sea-level change
through time and reliably compare instrumental compilations with proxy
reconstructions, it is necessary to account for the uncertainties that
characterize each data set. Previous studies used simple linear
regression models (most commonly polynomial regression), resulting in
overly precise rate estimates. We develop models to estimate rates of
sea-level change and account for all available sources of uncertainty
in instrumental and proxy-reconstruction data. Our response variable
with the proxy measurements is sea level after correction for GIA. Our
models place a Gaussian process (GP) prior on the rates of sea-level
change and the mean of the distribution assumed for the observed data
is the integral of this rate process. By embedding the integrated
Gaussian process (IGP) model in an errors-in-variables (EIV) framework
(which takes account of time uncertainty) and removing the estimate of
GIA, we quantify rates with better estimates of uncertainty than was
previously possible.

To demonstrate the application of these models, we apply them to an
example global tide-gauge data set [\citet{CandW2011}]. Our analysis of
this record indicates that the rate of GMSL rise increased continuously
since 1880~AD and the posterior estimate of the mean rate of sea-level
in 2009~AD is 1.92 mm$/$yr. The 95$\%$ credible interval for this mean is
1.84 to 2.03 mm$/$yr. We also apply the model to a late Holocene proxy
reconstruction from North Carolina [\citet{Kemp2011}]. Such
reconstructions are important to understand the response of sea level
to known climate variability such as the Medieval Climate Anomaly and
the Little Ice Age [\citet{Mann08}]. Application of our model to the
North Carolina proxy reconstruction indicates a posterior mean rate of
rise in this locality since the middle of the 19th century of 2.44
mm$/$yr. The 95$\%$ credible interval for this mean is 1.91 to 3.01
mm$/$yr. This result is in agreement with results from the tide-gauge
analysis and illustrates that the current rate of sea level is
unprecedented in at least the last 2000 years. The two examples show
the importance and utility of the new models in estimating dynamic
rates of sea-level change with full and formal consideration of the
uncertainties that characterize instrumental and proxy data sets.

\section{Sea-level data sets}\label{sec2}
This section describes how the global tide-gauge record [\citet
{CandW2011}] was compiled and how RSL in North Carolina was
reconstructed using proxies preserved in cores of coastal sediment
[\citet{Kemp2011}]. Although the methods for data collection are
specific to our case studies, the resulting records are typical of
available sea-level data sets.

\subsection{Tide gauges}\label{sec2.1}
\begin{figure}

\includegraphics{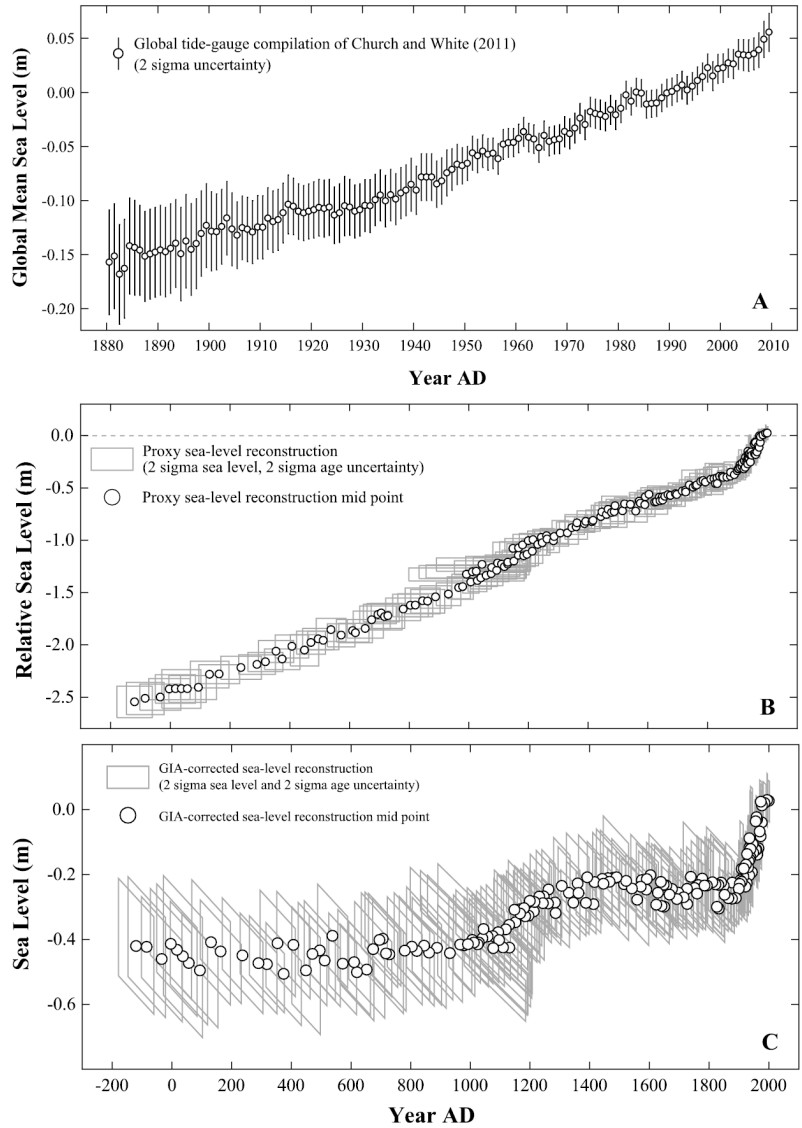}

\caption{\textup{(A)} Global tide-gauge record of \citet{CandW2011}. These
global mean sea-level data are
a compilation of individual tide-gauge records from sites located
around the world that
were individually corrected for the contribution of GIA. The data set
is characterized by vertical
(sea level) uncertainties (2 sigma uncertainty bands approximate the
95\% confidence interval),
but ages are fixed and known. \textup{(B)} Proxy reconstruction of RSL from
North Carolina, USA [\citet{Kemp2011}].
Individual data points (represented by rectangular boxes that
illustrate the 95\% confidence region)
are unevenly distributed through time and include age and sea-level
uncertainties. \textup{(C)} The North Carolina reconstruction following
correction for GIA.}
\label{fig1}
\end{figure}

Tide gauges are instruments that measure RSL multiple times each day at
a fixed coastal location. Monthly RSL averages for individual locations
are held by the Permanent Service for Mean Sea Level [Woodworth and Player
(\citeyear{Player03})].
The distribution of these locations is very uneven in time and space.
To reliably estimate rates and trends against a background of annual to
decadal variability, analysis of individual tide-gauge records is
commonly restricted to locations with more than $\sim$60 years of data
[\citet{Douglas01}]. GMSL is estimated by spatially averaging tide-gauge
records after individual records (irrespective of record length) were
corrected for GIA. The most commonly used data set is that of \citet
{CandW2011}, which includes annual sea-level data between 1880~AD and
2009~AD from up to 235 individual locations [Figure~\ref{fig1}(A)]. This data set
employed the spatial variability in sea level observed by satellites to
interpolate between tide-gauge locations and estimate global sea level.
Other studies produced alternative estimates of GMSL using different
methodologies to correct for GIA and to account for the uneven
distribution of tide gauges in time and space [e.g., \citet{Hay2015},
\citet{JJ2008}]. However, each of these compilations shared the basic
attributes of the Church and White (\citeyear{CandW2011}) data set in having a fixed
and known age, but estimated GMSL with uncertainty. Therefore, our
choice of example data set is typical of tide-gauge data, but our model
could also be appropriately applied to similar data sets and
potentially yield different estimates of rates of historical sea-level change.

\subsection{Salt-marsh reconstructions}\label{sec2.2}
Salt marshes keep pace with sea-level rise by accumulating sediment
[\citet{Morris02}]. As a result, modern salt marshes may be underlain by
several meters of sediment, which is an archive of past sea-level
changes. Cores are used to recover this coastal sediment for analysis.
The ages of discrete depths in the core are estimated using techniques
such as radiocarbon dating to provide a history of sediment
accumulation. Radiocarbon dates are calibrated into calendar ages and
assimilated with other chronological constraints (e.g., pollution
markers of known age) using an age-depth model.

For the North Carolina reconstruction [Figure~\ref{fig1}(B)], ages for the RSL
data were calculated from Bchron [\citet{bchron},
\citeauthor{Parnell08} (\citeyear{Parnell08,Parnell2011})],
a Bayesian, statistical age-depth model that estimates uncertain
interpolated ages between radiocarbon dated levels. This tool is
particularly useful in reconstructing RSL from a core of coastal
sediment, because most levels in the core were not directly dated.
Bchron assumes that the calibrated radiocarbon ages arise as
realizations of a Compound Poisson--Gamma (CPG) process, which enforces
the geological law of superposition. Bchron calibrates the radiocarbon
dates, estimates the parameters of the CPG and
identifies outliers. The ages and 1 sigma age errors used in the North
Carolina proxy reconstruction are the Bchron marginal means and
standard deviations for each layer in the core that was used to
reconstruct RSL, which we approximate as being normally distributed.
This would be a poor assumption for individual calibrated radiocarbon
dates that are skewed and multi-modal. However, the CPG produces
slightly more regular ages, and the effect is further reduced when
combined with our smoothing approach.

Core sediment contains the preserved remains of microorganisms such as
foraminifera. The distribution of foraminifera is controlled by tidal
elevation (i.e., sea level) because some species are more tolerant of
submergence by the tides than others [\citet{Scott78}]. The modern,
observable relationship between counts of foraminifera and sea level
provides an analogy for interpreting similar assemblages preserved in
core material. This analogy is exploited to reconstruct RSL using a
transfer function [\citet{Kemp13,Birks95,Horton06,Juggins2012}]. The
calibration of counts of foraminifera into estimates of RSL (via these
transfer functions) requires further statistical modeling techniques
that we do not discuss here. The transfer function output returns an
estimate of the error associated with each fossil sample. This is given
by the root mean square error of prediction of a training set, derived
using a separate test set, or by internal cross-validation. We include
these error estimates, assumed to be 1 sigma uncertainties, as an input
to our model. The validity of this approach was demonstrated by
comparison between reconstructions and instrumental measurements from
nearby tide gauges [e.g., \citet{KandH09}]. To extract climate-driven
rates of sea-level rise, the RSL reconstructions are corrected for GIA,
which over the last 2000 years is assumed to be a constant rate because
of the slow response time of the solid Earth [\citet{Peltier04}]. The
GIA corrected reconstruction for North Carolina is shown in Figure~\ref{fig1}(C).

\section{Previous work}\label{sec3}
In this section we review how rates of sea-level change are estimated
from uncertain data in existing literature. We also describe the
stochastic methods that we employed in this paper.

\subsection{Sea-level rise: Rates and accelerations}\label{sec3.1}
The motivation for analyzing tide-gauge records and reconstructing RSL
is to establish how unusual modern rates of sea-level rise are in
comparison to longer term trends and for understanding the role of
climate variability as a driver of sea-level change [e.g., \citet
{Donnelly04,Engelhart09,Shennan02}]. Comparisons of past and present
rates are only complete and fair if all sources of uncertainty are
accounted for. The global tide-gauge record is the primary source of
historic and current sea-level data. The record includes sea-level
uncertainty that is greater earlier in the record because it is based
on fewer individual records that are unevenly distributed in space with
a bias toward western Europe and North America [\citet{JJ2008}]. The age
of each annual sea-level observation is fixed and known. Tide-gauge
records are commonly analyzed using simple linear regression to
estimate a rate of sea-level rise for the entire record or a shorter
segment [e.g., \citet
{Barnett84,NewJJ,Douglas01,Gornitz82,Peltier91,Sallenger2012,CandW2006,Holgate04}].
For example, \citet{CandW2011}
calculated the mean rate of global sea-level rise to be 1.6 mm$/$yr $\pm$
0.3 mm from 1880~AD to 2009~AD compared to 1.1 mm$/$yr $\pm$ 0.7 mm between
1880~AD and 1936~AD, and 1.8 mm$/$yr $\pm$ 0.3 mm after 1936~AD. Satellite
altimetry data have also been analyzed in this way to estimate a rate
of GMSL rise of 3.4 mm$/$yr $\pm$ 0.4 mm between 1993~AD and 2008~AD [\citet
{Nerem2010}].

Similar approaches were widely employed to characterize acceleration or
deceleration of sea-level rise, where a polynomial rather than linear
function was fitted to the tide-gauge record [e.g., \citet
{Boon2012,HoustonandDean,JJ2008,Woodworth09}]. For example, \citet{CandW2011}
estimated a sea-level acceleration of 0.009 mm$/$yr$^2$ $\pm$ 0.003
mm$/$yr$^2$ for the period 1880~AD to 2009~AD.
In contrast, \citet{HoustonandDean}
obtained a small sea-level deceleration ($-0.0123 \pm 0.0104$
mm$/$yr$^2$) by selectively analyzing U.S. tide gauges from 1930~AD to
2010~AD and suggested similar decelerations for the global data set over
the same time interval. A limitation of subdividing the tide-gauge
record into segments identified by visual inspection is that individual
data points are ascribed undue importance and information is lost in
the autocorrelated data set by discarding earlier and/or later
intervals. Furthermore, the estimated rates of change are sensitive to
the data included, making comparisons among data sets difficult. For
example, the rate of sea-level change measured by satellite altimetry
since 1993~AD is greater than the ``current'' (1936--2009~AD) rate often
quoted from Church and White's (\citeyear{CandW2011}) analysis of their global
tide-gauge record. Although the estimated rates of change are
different, Church and White emphasized the agreement between the two
methods of measuring sea-level change over the period where both data
sources are available.

In considering proxy reconstructions with bivariate uncertainties, some
studies divided the data series into sections based on changes in slope
that were qualitatively positioned by the researcher at a single time
point [e.g., \citet{Gehrels2012}]. Consequently, a rate of change was
calculated for each segment of the sea-level reconstruction by simple
linear regression of midpoints with no formal consideration of age and
sea-level uncertainty or their covariance. Other studies used an EIV
change point approach to objectively place changes in slope across a
range of timings and to estimate linear rates for each segment with
consideration of uncertainty [\citeauthor{Kemp2011}
(\citeyear{Kemp2011,Kemp13}), \citet{IOW2014}]. A
limitation of this approach is that phases of persistent sea-level
behavior are approximated by linear trends that do not accurately
represent the underlying physics of sea-level change and mask (to some
degree) the continuous evolution of sea level through time.

\subsection{Stochastic processes and rate estimation}\label{sec3.2}
The model we propose makes use of the EIV approach, where we do not
assume that the explanatory variable (which we denote as $x$) is known,
but that it is instead measured with some error [\citet{EIV}]. The EIV
approach can be used with multivariate and hierarchical models
including our application to proxy sea-level reconstructions with age
and sea-level errors. We embed our EIV regression within a
nonparametric model.

We use a GP as a prior on the rate process, which is then integrated to
estimate sea level. The opposite approach, where a GP is placed on the
data itself and then differentiated to produce rates, has a long
literature [\citet{ohagan1992,Cramer1967}], both where the derivatives
were observed and where they were estimated. A fuller description of
GPs is found in \citet{Rasmussen96} and \citet{Rasmussen06GP}. Most
recently in sea-level research, \citet{kopp2013}
employed
an empirical
Bayesian analysis that used GPs to assess the statistical significance
of the ``hot spot'' of sea-level acceleration in the mid-Atlantic and
northeastern regions of the United States.
Hay et al. (\citeyear{Hay2015}) use GPs for rate estimation and to assess the robustness of their probabilistic reanalysis of GMSL.
However, GPs are not the
only means for creating rate estimates. Other work exists in the field
of splines [e.g., \citet{Mardia94,Bsplines}] or in diffusion
processes and differential equation models [e.g., \citet{Hua08}].

We do not cover spatio-temporal modeling of sea-level rates in this
paper, focusing instead on individual sites. The behavior of sea level
in space is highly irregular and relates to numerous physical features
and processes that are beyond the range of the statistical models we
discuss. We focus on a novel EIV-IGP approach. The GP has advantages
over other methods mentioned previously due to its simplicity and
flexibility despite using only a small number of parameters. The IGP we
employ is an inverse model where the GP is applied to the rate process
rather than the observed data. \citet{Holsclaw2013} outline a method
for posterior computation of such models which we employ in the next section.

\section{Methods}\label{sec4}
In this section we outline the EIV-IGP model used to estimate past sea
level while accounting for age uncertainty. We apply this model to the
North Carolina proxy reconstruction in Section~\ref{sec5}. Our first case study
(the global tide-gauge record) requires a slightly simplified version
of this model (which we term S-IGP), because the data has fixed and
known ages and, therefore, lacks age uncertainity. The raw data are
scalars $(y_i, \sigma_{y_i},x_i, \sigma_{x_i})$ for $i = 1,\ldots,n$ data
points, where $y_i$ is the RSL measurement and $\sigma_{y_i}$ is the
sample-specific estimate of uncertainty for the measurement which is
one standard deviation, $x_i$ is the estimated age measurement from the
chronology model, and $\sigma_{x_i}$ is the age standard deviation,
also taken from the chronology model. Ignoring GIA correction for the
moment, we can write
%
\begin{eqnarray}
y_i &=&\alpha+ h(\chi_i) +\varepsilon_i,
\qquad i=1, \ldots, N,
\\
x_i &=&\chi_i + \delta_i,\qquad i=1,
\ldots, N, \label{xi}
\end{eqnarray}
where the errors $\varepsilon_i \sim N(0, \sigma_{y_i}^2 + \tau^2)$ are
independent and $\tau^2$ is a micro-scale variance term. Modelers
sometimes separate the micro-scale variation using $\eta$ which
captures the micro-scale variation and $\varepsilon$ which captures the
pure measurement error [\citet{Banerjee2012}]. The methods used to
reconstruct sea level, described in Section~\ref{sec2.2}, assume that the
distribution of microorganisms such as foraminifera is controlled by
tidal elevation (i.e., sea level). However, foraminifera abundances are
also affected by other sources of noise [e.g., influence of additional
environmental variables such as salinity or sediment texture; \citet
{Horton1999}]. As a result, it is necessary to include $\tau^2$ in the
model to account for any unexplained variation that may be present in
the data. $\delta_i \sim N(0, \sigma_{x_i}^2)$, $\alpha$~is a constant
intercept parameter, $h(\chi)$ is a stochastic process in continuous
time that represents the underlying evolution of RSL and $\chi_i$ is
the true unobserved age for observation $i$. The mean of the
distribution for the observed data is dependent on the stochastic
process that we want to estimate and the model is set up to have a
classical EIV structure. The key parameters are those in $h$ and the
micro-scale variance $\tau^2$. The estimated true ages $\chi_i$ are
nuisance parameters. Our focus lies in posterior inference about $h$
and, most importantly, its derivative.

As discussed in Section~\ref{sec3.2}, there are numerous nonparametric priors on
functions that provide stochastic derivatives, though our situation is
complicated by the inclusion of age uncertainties. If the data were
modeled with a GP, we would write $y_i=\alpha+g(\chi_i)+\varepsilon_i$,
where $g(\chi_i)$ is a GP with a mean function $\mu_g$ (which we set to
0) and a covariance function denoted $\upsilon^2C_g(\chi_i,\chi_j)$.
One approach to obtaining derivatives would be to fit the GP to the
data and differentiate the correlation function. However, \citet
{Holsclaw2013} outline how this approach can be inadequate due to loss
of information when differentiating. They state that, if we fit the
data process directly and then differentiate, we lose information about
the derivative process, as, when observational error is present, the
prediction of the derivative process is degraded [\citet{Stein1999}].
Since our focus is directly on the rate process, $g'(\chi_i)=w(\chi
_i)$, we prefer to place a GP prior distribution on this and integrate
to create estimates of the mean of the observed data, which we now
denote $h$.

Writing $h(\chi) = \int_0^{\chi} w(u)\,du$, we place a GP prior on $w$.
The integration limits were simplified to start at 0 by readjusting $x$
in the model setup. For our model we chose a GP with a mean function
$\mu_w$ (which we set to 0) and a stationary powered-exponential
covariance function, which we denote $C_w(\chi_i,\chi_j)$. The
exponential covariance function is appropriate due to the underlying
smooth nature of the sea-level data [abrupt changes in sea level are
unlikely except in areas of instantaneous tectonic deformation such as
those caused by megathrust earthquakes; e.g., \citet{Atwater87}]. We use
a re-parameterized version so that $C_w(\chi_i,\chi_j) ={ \rho^{|\chi
_i-\chi_j|}}^{\kappa}$ with $\rho\in(0,1)$ and $\kappa\in(0,2]$.
The distribution of the observed data process $h$ is available also as
a GP, where the covariance function for $h$ can be obtained by
integrating the covariance function for the derivative process ($w$)
twice. The resulting covariance of the IGP is $K_h(\chi_i,\chi_j) =\int_0^{\chi_i} \int_0^{\chi_j} C_w(u, v) \,du \,dv$ and $h$ created from such
a situation will be nonstationary, which will allow for the smoothness
of the function to vary with repect to the input space, resulting in a
more flexible model.

A solution to the problematic double integration is provided by \citet
{Holsclaw2013}. They used an approach given in \citet{Yaglom2011} to
bypass the calculation of the double integral by approximating the
integrated process on a grid $x^* = (x_1^*\cdots x_m^* )$ for arbitrarily
large $m$. This yields
%
\begin{equation}
h(\chi) \approx K_{hw}^*C^{{**}^{-1}}_w
w_m,
\end{equation}
where $C^{**}_w = [C_w(x_i^*, x_j^*)]_{i,j=1}^m$ is an $m \times m$
matrix containing the covariance function for the derivative process
and $w_m \sim \operatorname{GP}(\mu_w, \upsilon^2C^{**}_w)$. Most usefully, $K^*_{hw}$
is the covariance between the rate process and the integrated process,
and so only involves integrating the covariance once, that is, $
K_{hw}^* (\chi_i,x^*_j) = \int_0^{\chi_i} C_w(u,x_j^*) \,du$. This
integral is calculated numerically, using Chebyshev--Gauss quadrature
[\citet{quadrature}].

The \citet{Holsclaw2013} approach replaces the integral estimate with
the conditional mean of the integrated process given the derivative
process, while ignoring any conditional variance. This approach is
strongly related to that of the predictive processes [PP: \citet
{Banerjee2008}]. In the PP approach, a spatial covariance matrix is
approximated onto a smaller grid also by its conditional mean,
resulting in smaller matrix manipulations for large spatial problems.
However, a significant disadvantage of the PP is that the low rank
approximation can yield poor estimates of the correlation structure. By
contrast, our processes are one dimensional and we can set the grid
size $m$ large to arbitrarily reduce the approximation error with
little computational cost. In that sense it has elements in common with
high-rank approximations such as \citet{Lindgren2011}.

\begin{figure}[b]

\includegraphics{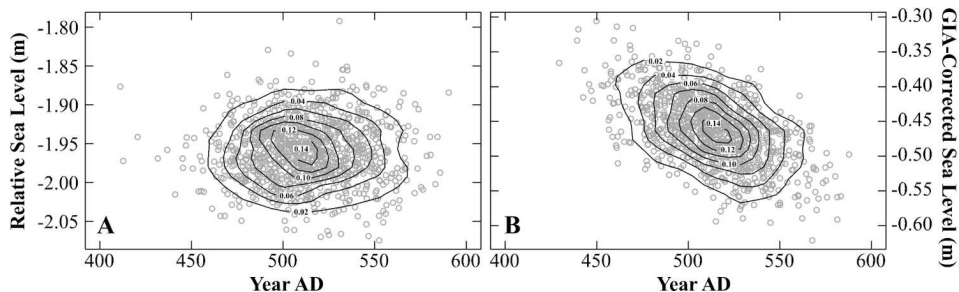}

\caption{Example of correcting a single data point in the North
Carolina proxy sea-level reconstruction for the effect of GIA. N=1000
data points were simulated from the bivariate distribution of the RSL
reconstruction before \textup{(A)} and after \textup{(B)} it was corrected for GIA. Since
GIA is a rate (in~mm$/$yr), this correction results in bivariate
correlated errors.}\label{fig2}
\end{figure}

Last, we must account for GIA in our model for the proxy measurements.
This introduces an extra fixed parameter $\gamma$ (measured here in
mm$/$yr) to account for land-level movements at an individual site. The
GIA correction involves subtracting $x_i$ from the year of core collection,
denoted $t_0$. This is then multiplied by the rate of GIA $\gamma$ and
added to $y_i$ for each observation $i$. The introduction of the GIA
parameter raises or lowers the sea level associated with each data
point, and additionally introduces a correlation between the age and
sea-level reconstructions since older sea-level observations are
raised/lowered to a greater degree. As an illustration, consider the
single example from the North Carolina proxy reconstruction shown in
Figure~\ref{fig2}(A). The data point is given by the quadruple $(y_i,x_i,\sigma
_{x_i},\sigma_{y_i})$ with the density of this data point represented
as contours, and samples shown for illustration. Once the GIA effect is
removed, we obtain Figure~\ref{fig2}(B), where the left-hand side of the density
has been raised to a greater degree than the right-hand side because it
is older.

Algebraically, the GIA effect can be removed via an affine
transformation of the data and the variance matrix by matrices $A= \bigl[
{1 \atop  -\gamma }\enskip{0\atop  1}\bigr]$ and $b = \bigl[
{ 0 \atop \gamma t_0}
 \bigr]$. The GIA-corrected model is now
%
\begin{equation}
Az_i+b \sim N \bigl(\mu_i,AV_iA^T
+ \tau^2 B \bigr),\qquad i=1, \ldots, N, \label{modeleqn}
\end{equation}
where $z_i = \bigl[
 {x_i \atop y_i}
 \bigr]$, $\mu_i = \bigl[
{\chi_i \atop \alpha+h(\chi_i)}
 \bigr]$, $V_i = \bigl[
{ \sigma^2_{x_i} \atop 0} \enskip{ 0 \atop \sigma^2_{y_i}}
 \bigr]$ and $B = \bigl[
{ 0 \atop 0} \enskip {0 \atop 1}
 \bigr]$. Since $Az_i$ and $AV_iA^T$ are both deterministic functions
of the data, they can be calculated off-line prior to any analysis.

The rate of GIA to be applied is spatially variable because of the
underlying physical process [\citet{Engelhart09}]. For our North
Carolina case study, where there are two sites, we apply the rates of
0.9 mm$/$yr and 1 mm$/$yr that were used in the original publication.
Equation \eqref{modeleqn} forms the likelihood for the observed data
based on the EIV-IGP model. This completes our model specification.

All the models we outline were fitted in the JAGS (Just Another Gibbs
Sampler) language [\citet{JAGS}]. JAGS is a tool for analysis of
Bayesian hierarchical models using Markov Chain Monte Carlo (MCMC)
simulation. Although writing customized MCMC sampling algorithms can in
some cases be relatively straightforward, it has become more common
practice to make use of Bayesian MCMC fitting software such as the
Bayesian analysis Using Gibbs Sampling (BUGS) software. JAGS is an
engine for running BUGS and allows users to write their own functions,
distributions and samplers. JAGS offers cross-platform support and a
direct interface to R using the package rjags [\citet{rjags}].

We validated our model using two methods. First, we simulated data
under ideal and nonideal conditions. The ideal scenario is one where
the parameters are simulated from the same distributions as the priors
that are placed on the parameters. The nonideal scenarios lead to the
prior distributions over/underestimating the mean and the variance of
the parameters. The aim was to determine, for each scenario, the
coverage probabilities for the true rate process within 95\% and 68\%
credible intervals. Second, we performed a 10-fold cross-validation on
our case study data. Results were highly satisfactory for both
validation methods and we are confident that using this model for
instrumental and proxy sea-level data allows us to estimate the
underlying rates of sea-level change with a high degree of accuracy.
Further details of how the validation was carried out along with
results can be found in the \hyperref[app]{Appendix}. All code is available in the
supplementary materials [\citet{supplement}].

\section{Case studies}\label{sec5}
In this section we outline our prior distributions in further detail
for each of our case studies. In the first case study we use tide-gauge
measurements which have small age uncertainties and so are ignored,
effectively removing the EIV structure and allowing us to demonstrate
the IGP aspect of the model. Our second case study, the proxy data,
contains all the elements outlined in this section. To illustrate the
utility of the S-IGP and EIV-IGP models, we apply them to the global
tide-gauge record since 1880~AD [\citet{CandW2011}] and a proxy RSL
reconstruction spanning the last 2100 years [\citet{Kemp2011}]. The goal
is to obtain the posterior distribution of sea level and of the rate
process of interest.

For both case studies we initially ran the appropriate model for 5000
iterations with a burn-in of 500 that we thinned by 3. In both cases we
saw good convergence. We then ran the model for a long run of 50,000
iterations to ensure convergence remained and results were consistent.
The R package coda [\citet{coda}] was used to run diagnostics. We used
autocorrelation plots, Geweke plots [\citet{Geweke92}], the Gelman and
Rubin diagnostic [\citet{GandR92}] and the Heidelberger and Welch
diagnostic [\citet{heidel}], which all indicated model convergence. We
also ran multiple chains from different starting values to ensure good mixing.

\subsection{Global tide-gauge record}\label{sec5.1}
A complete description of the approach and methods employed to generate
this data set is presented in Church and White (\citeyear{CandW2006,CandW2011}).
The data file includes 3 columns: time in years AD, GMSL in meters, and
a one-sigma sea-level error in meters.

The Simple Integrated Gaussian Process (S-IGP) model was used to
analyze this data set. The distribution for the observed data is
%
\begin{eqnarray}
y_i &\sim& N \bigl(\alpha+h(x_i),\sigma_{y_i}^2+
\tau^2 \bigr),
\\
h(x)& \approx& K_{hw}^*C^{{**}^{-1}}_w w_m,
\end{eqnarray}
where $h(x)$ is the approximation to the IGP described in Section~\ref{sec4}.
$C^{**}_w ={ \rho^{|x^*_i-x^*_j|}}^{\kappa}$ is an $m \times m$ matrix
containing the covariance function for the derivative process and $w_m
\sim \operatorname{GP}(\mu_w, \upsilon^2C^{**}_w)$. Recall, $K^*_{hw}$ is the
covariance between the rate process and the integrated process, that
is, $ K_{hw}^* (x_i,x^*_j) = \int_0^{x_i} C_w(u,x_j^*) \,du$.

Prior distributions were specified for each unknown parameter. The
correlation parameter $\rho$ was defined on the interval $(0,1)$. The
tide-gauge record [\citet{CandW2011}] spans a relatively short period of
time, during which there was a single mode of climate warming and
sea-level rise [\citet{Rahmstorf07}]. So even though this record is
highly correlated, climate forcing, as opposed to time change, is the
driver for sea-level change over this instrumental period. Therefore,
we set a mildly informative prior for $\rho$ that favors low values of
the correlation parameter that are close to 0.2, where $p(\rho
)=\operatorname{Beta}(2,8)$. Another somewhat informative prior was used for $\tau^2$.
To determine a prior for this parameter, we considered other global
tide-gauge compilations such as \citet{JJ2008}. The data supplied for
this record have associated standard errors for each sea-level
measurement. Details of how these errors are determined can be found in
\citet{JJ06}. These standard errors range from 0.01--0.07 m. In choosing
our prior we used this information, but we do not restrict $\tau$ to be
within this range, instead we chose to conservatively place a prior on
$\tau^2$ that favors values for $\tau$ close to 0.1 m, where $\tau
^2\sim \operatorname{Gamma}(0.1,10)$.

We decided on a prior for $\upsilon^2$, the variance of the rate
process, by looking at the information currently available regarding
the rate of global sea-level rise. Between 1950~AD and 2000~AD trends in
global average rates of sea-level rise varied from 0 to 4 mm$/$yr [\citet
{CandW05}]. Over multi-centennial timescales during the last 2000 years
(prior to industrialization), global sea level was likely close to
stable after correction for land-level movements (i.e., rate $\sim$0~mm$/$yr). Alternatively, at decadal to multi-decadal time scales, higher
regional rates (up to 4 mm$/$yr) are observed in instrumental records
after correction for land-level movements. A GP prior centered on 0 was
used to describe the rate process for our model. The prior information
suggests that rates can reach up to 4~mm$/$yr. Therefore, we deemed the
range of the rate of sea level, $-4$ to 4 mm$/$yr, appropriate. If this
range is treated as a 95\% confidence interval, it is reasonable to
assume that the standard deviation is $\sim$ 2 mm$/$yr. Hence, we set up
the prior for $\upsilon^2$ to favor values close to 4, where $\upsilon
^2\sim \operatorname{Gamma}(80,20)$. An uninformative normal prior is placed on the
unknown intercept parameter~$\alpha$.

\begin{figure}

\includegraphics{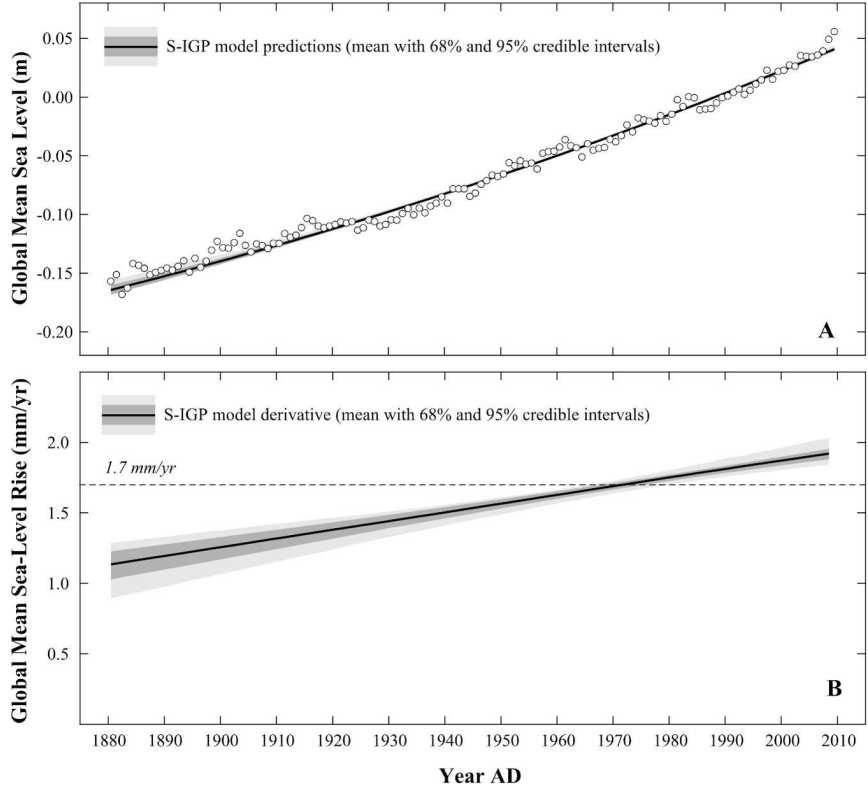}

\caption{\textup{(A)} Predictions for GMSL since 1880~AD generated by fitting the
S-IGP model to the instrumental data set. Shading denotes 68\% and 95\%
credible intervals for the posterior mean fit. \textup{(B)}~Rate of global
sea-level rise calculated as the derivative of the fitted model.
Shading denotes 68\% and 95\% credible intervals for the posterior mean
of the rate process.} \label{fig3}
\end{figure}

The analysis of the global tide-gauge record is presented in Figure~\ref{fig3},
which shows the GMSL predictions estimated from our model (A) and our
rate estimates~(B). The rate of GMSL rise estimated from a linear
regression analysis of the global tide-gauge record for the entire
period 1880~AD to 2009~AD was 1.5~mm$/$yr [\citet{CandW2006}]. This is
consistent with the average rate suggested in Figure~\ref{fig3}(B). The rate of
sea-level rise from 1900~AD to 2009~AD was 1.7 mm$/$yr $\pm$ 0.3 mm [\citet
{CandW2006}]. The S-IGP model indicates that this rate occurred from
approximately 1965~AD to 1975~AD. Furthermore, the model indicates that
the rate of GMSL rise actually increased (accelerated) constantly
through time from 1.13 mm$/$yr in 1880~AD to 1.92 mm$/$yr in 2009~AD
[Figure~\ref{fig3}(B)]. The recognition of accelerating sea-level rise agrees with
projections for the 21st century that can only be realized with
continued acceleration [\citet{IPCC}]. We demonstrate that the S-IGP
model negates the need to analyze specific intervals of temporal data
and consequently provides more accurate and representative estimates of
the constantly evolving rate of sea-level change.

\subsection{North Carolina proxy reconstruction}\label{sec5.2}
The example data set from North Carolina is a proxy reconstruction
spanning the last $\sim$2100 years that was developed from cores of
salt-marsh sediment located at two sites (Tump Point, 34$^{\circ
}58'12''$N 76$^{\circ}22'48''$W; and Sand Point, 35$^{\circ}53'05''$N
75$^{\circ}40'51''$W) that are 120~km apart [\citet{Kemp2011}]. As such,
it provides a regional record of RSL change for North Carolina. The
correction for GIA was estimated from a regional database of late
Holocene relative sea-level reconstructions [\citet{Engelhart09}]. The
rate of GIA is 0.9 mm$/$yr at Tump Point and 1.0 mm$/$yr at Sand Point. The
data file includes 4 columns: RSL in meters, age in year AD, a~one-sigma RSL error, and a two-sigma age error.

\begin{figure}

\includegraphics{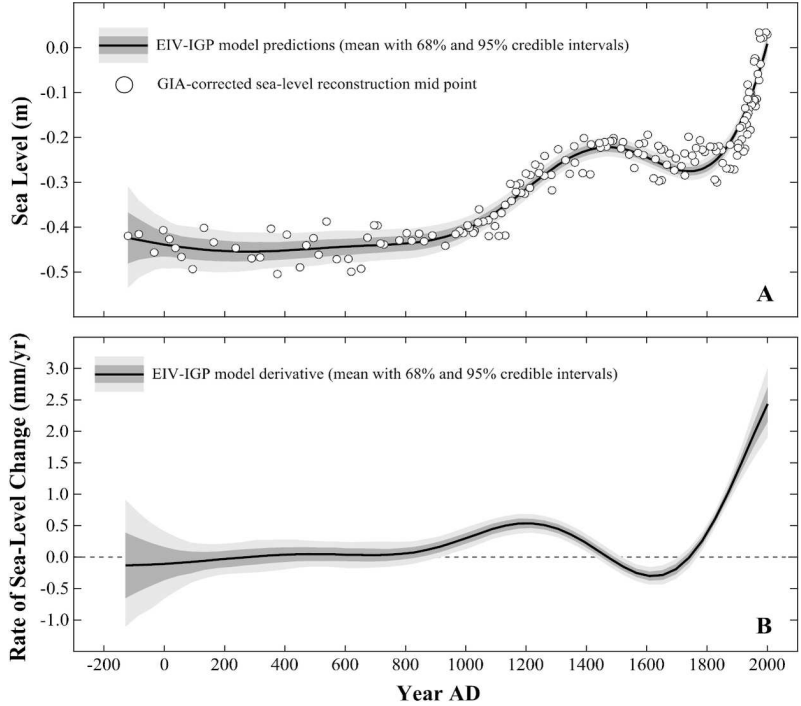}

\caption{\textup{(A)} Predictions for North Carolina sea level generated by
fitting the EIV-IGP model. Shading denotes 68\% and 95\% credible
intervals for the posterior mean fit. \textup{(B)} Rate of sea-level change in
North Carolina calculated as the derivative of the fitted model.
Shading denotes 68\% and 95\% credible intervals for the posterior mean
of the rate process.}\label{fig4}
\end{figure}

The EIV-IGP model, described in detail in Section~\ref{sec4}, was used to
analyze this data set. Prior distributions were specified for each
unknown parameter. As with the S-IGP model, the correlation parameter
$\rho$ is defined on the interval $(0,1)$. The chosen prior $p(\rho
)=\operatorname{Beta}(2,8)$, which suggests a mean of approximately 0.2 with a
standard deviation of approximately 0.1. This assumes that data points
more than 1000 years apart have minimal effect on one another. This is
a reasonable assumption given that the reconstruction spans a 2100 year
time period and includes multiple phases of sea level and climate
behavior, including the warmer Medieval Climate Anomaly, cooler Little
Ice Age, and very warm 20th and 21st centuries [\citet{Mann08}]. We used
the same prior for the variance parameter $\tau^2$ as for the previous
case study. Following the same reasoning as with the tide-gauge data in
Section~\ref{sec5.1}, a gamma prior, $\upsilon^2\sim \operatorname{Gamma}(80,20)$, was used for
the variance of the derivative process. An uninformative normal prior
was placed on the unknown intercept parameter $\alpha$.

Application of the EIV-IGP model to the proxy sea-level reconstruction
from North Carolina shows four persistent phases of sea-level behavior
[Figure~\ref{fig4}(A)]. The model predictions are a good fit to the proxy
reconstructed data which gives confidence in the model. From the start
of the record at approximately 100~BC to 1000~AD there is little change
in sea level following correction for GIA. The period from 1000~AD to
1400~AD is characterized by sea-level rise. Between 1400~AD and about
1850~AD there was a fall in sea level and since 1850~AD sea level rose
rapidly in North Carolina. This evolution in sea level is reflected in
the modeled rate of sea-level rise [Figure~\ref{fig4}(B)], where the first period
has a mean sea-level change of approximately 0 mm$/$yr. The second period
saw a maximum rate of rise reach a posterior mean value of 0.53 mm$/$yr
with a 95\% credible interval for this mean of 0.39 to 0.68 mm$/$yr,
which \citet{Kemp2011} attributed to a warmer climate during the
Medieval Climate Anomaly. The sea-level fall between 1400~AD and 1850~AD
occurred at a maximum rate of 0.3 mm$/$yr with a 95\% credible interval
for this mean of 0.16 to 0.43 mm$/$yr and was likely a sea-level response
to the cooler Little Ice Age [\citet{Kemp2011}]. The transition from the
Little Ice Age is marked by a dramatic increase in the rate of
sea-level rise that continues to a mean rate of 2.44 mm$/$yr in 2000~AD
with a 95\% credible interval of 1.91 to 3.01 mm$/$yr. The rate of
sea-level rise since the middle of the 19th century is without
precedent in North Carolina for at least the previous 2000 years. The
modeled mean rate of rise departs from earlier 95\% credible intervals
at around 1845~AD.

\section{Conclusion}\label{sec6}
Taking into account all sources of uncertainty (temporal and vertical)
when estimating sea-level trends is essential to allow instrumental
measurements and proxy reconstructions of sea level to be compared
directly and fairly. Previous analysis incorrectly ignored some or all
of the uncertainties. We proposed and validated a model that allows for
the direct estimation of rates of sea-level change while quantifying
uncertainties more thoroughly than previously possible. The method
involves a nonparametric reconstruction of the derivative process. A GP
prior is placed on the derivative process and we view the mean of the
distribution assumed for the observed data to be the integral of this
process. For our case study data, the derivative at a particular time
point is representative of the rate of sea-level change at that time
point. This enables us to estimate instantaneous rates of change and
observe the constant evolution of dynamic sea-level rise through time.
The model also provides a flexible fit and allows us to estimate the
uncertainty about the rate process of interest.

Our analysis of
the global tide-gauge record shows that the rate of GMSL rise increased
(accelerated) continuously from 1.13 mm$/$yr in 1880~AD to 1.92 mm$/$yr in
2009~AD. Application of our model to an example proxy sea-level
reconstruction from North Carolina quantified the changing rate of
sea-level rise through the Medieval Climate Anomaly, Little Ice Age and
20th century. The posterior mean rate of rise in North Carolina at
2000~AD was 2.44 mm$/$yr with a 95$\%$ credible interval of 1.91 to 3.01
mm$/$yr. This is the fastest rate of rise in the 2000-year long reconstruction.

\begin{appendix}\label{app}
\section*{Appendix: Model validation}
\subsection{Simulated scenarios}
In this section we demonstrate the validity of our model. Through the
use of simulated data, parameters $\alpha$ and $\tau^2$ (see
Section~\ref{sec4}), proved to be robust. Within reason, there was no difficulty in
estimating the values of these parameters, regardless of prior choice.
We found the parameters that related to the GP, that is, $\sigma_g^2$
and $\rho$, were the more sensitive parameters in the model and, as a
result, the validation focused on these. For the purposes of this
validation we used a simpler model, that is, the version that is not
set in the EIV framework. The parameters that were introduced in cases
where an errors-in variables approach was necessary were all estimated
directly from the data and, thus, we excluded this component of the
model in the validation process in order to simplify things. Therefore,
the data was simulated from the following distribution:
%
\begin{eqnarray}
y_i & \sim &N \bigl(\alpha+ h(x_i), \tau^2
\bigr),\qquad i=1,\ldots,N,
\\
h(x_i) &\approx& K_{hw}^*C^{{**}^{-1}}_w
w_m,
\end{eqnarray}
where $C^{**}_w = [C_w(x_i^*, x_j^*)]_{i,j=1}^m$ is an $m \times m$
matrix containing the covariance function for the derivative process
and $w_m \sim \operatorname{GP}(\mu_w, \upsilon^2C^{**}_w)$. $K^*_{hw}$ is the
covariance between the rate process and the integrated process.

To validate the model, we considered several different scenarios under
ideal and nonideal conditions. For each scenario we simulated values
for the unknown parameters, which in turn were used to simulate data
from an integrated GP model. Data simulation required simulation of the
underlying rate process, which, based on our model assumptions, is a
GP. Therefore, we knew the true underlying rate process. As the focus
of this work is in establishing rates of sea-level change, our primary
concern was whether or not our model was successful in estimating the
true underlying rate process. We observed how often the true rate falls
within the 95\% and 68\% credible intervals for the rate predicted from
the model.

For the purposes of this validation, the priors that were placed on the
parameters $\sigma_g^2$ and $\rho$ were $\sigma_g^2 \sim \operatorname{gamma}(10,10)$
and $\rho\sim \operatorname{beta}(2,8)$. Therefore, $\sigma_g^2$ will be centered
around 1 with a variance of 0.1 and $\rho$ will be centered around 0.2
with a variance of 0.01. In scenario (a) the parameter values came from
the same distributions as our priors. This was the ideal case and we
expected the model to perform best under these conditions. In scenarios
(b) and (c) we simulated the parameters so that our prior assumptions
were underestimating/overestimating the means, respectively. In
scenarios (d) and (e) we simulated parameter values so that our prior
assumptions were underestimating/overestimating the variances,
respectively. Finally, for scenarios (f) and (g) we simulated parameter
values so that our prior assumptions were
underestimating/overestimating both the mean and variances. 500
simulations were run for each scenario. The 95\% and 68\% coverage
probabilities were observed. An average over the 500 simulations was
taken for our validation results. The results of this validation are
shown in Table~\ref{ex:table}.

\begin{table}
\tabcolsep=0pt
\caption{Validation results for simulated scenarios \textup{(a)--(g)}}
\label{ex:table}
\begin{tabular*}{\textwidth}{@{\extracolsep{\fill}}ld{1.1}d{1.2}cd{1.3}cc@{}}
\hline
 & \multicolumn{2}{c}{\textbf{Simulated} $\bolds{\sigma^2_g}$}& \multicolumn
{2}{c}{\textbf{Simulated} $\bolds{\rho}$} & &\\[-6pt]
 & \multicolumn{2}{c}{\hrulefill}& \multicolumn
{2}{c}{\hrulefill} & &\\
\multicolumn{1}{@{}l}{\textbf{Scenario}} &
\multicolumn{1}{c}{\textbf{Mean}} & \multicolumn{1}{c}{\textbf{Var}} &
\multicolumn{1}{c}{\textbf{Mean}} & \multicolumn{1}{c}{\textbf{Var}} &
\multicolumn{1}{c}{\multirow{2}{85pt}[10pt]{\centering\textbf{Coverage probability (95\% CI)}}}&
\multicolumn{1}{c@{}}{\multirow{2}{85pt}[10pt]{\centering\textbf{Coverage probability (68\% CI)}}}\\
\hline
(a) (Ideal) &1&0.1&0.2&0.01&0.954&0.679\\
(b) &2&0.1&0.2&0.01&0.969&0.733\\
(c)&0.5&0.1&0.1&0.01&0.931&0.656\\
(d)&1&0.5&0.2&0.1&0.868&0.604\\
(e)&1&0.02&0.2&0.001&0.963&0.715\\
(f)&2&0.5&0.4&0.1&0.962&0.716\\
(g)&0.5&0.02&0.1&0.001&0.936&0.661\\
\hline
\end{tabular*}
\end{table}

The model was capable of estimating the rate process, even if the prior
distributions for the parameters were over/underestimating means and
variances. For the ideal scenario the true rate fell within the 95\%
credible interval and 68\% credible interval of the estimated rate
approximately 95\% and 68\% of the time as expected. For scenarios (b)
and (e) the rate fell into the credible intervals a higher proportion
of the time. This suggests that underestimating the mean values of our
parameters or overestimating the variance of our parameters will result
in wider than expected credible intervals for the rate. For scenarios
(c) and (d) the rate fell into the credible intervals a lower
proportion of the time. This suggests that overestimating the mean
values of our parameters or underestimating the variance of our
parameters will result in narrower credible intervals than expected for
the rate.

For scenario (f) the rate fell into the 95\% and 68\% credible
intervals more than 95\% and 68\% percent of the time. For this
scenario the priors were underestimating both the mean and variance for
the parameters of interest. From the cases where the mean and variance
were underestimated separately [i.e., (b) and (d)], the results were
wider credible intervals and narrower credible intervals, respectively.
Comparing these results with case (f), it appears that underestimating
the mean dictated the results and caused the credible interval for the
rate to be wider than expected. For scenario (g) the true rate falls
into the 95\% and 68\% credible interval less than 95\% and 68\% of the
time. In this case, when the means and variances were overestimated
[i.e., (c) and (e)], the results indicated narrower and wider credible
intervals, respectively. When compared with scenario (g), this
suggested that overestimating the means dictated the results and caused
the credible intervals to be narrower than expected.

From this validation we are made aware of some of the consequences of
misspecifying prior distributions for the parameters $\sigma_g^2$ and
$\rho$ in our model. The results indicated that although for all of the
cases excluding the ideal scenario we overestimate or underestimate our
confidence around the rate, we do not over/underestimate the credible
intervals by enough to cause concern.

\subsection{10-fold cross-validation}
A second method we used to validate our model was a 10-fold
cross-validation. We performed this on both our case study data sets.
Each observation was numbered $1:N$, where $N$ was the total number of
observations. A random permutation of these numbers was taken using a
function in R. The first 10\% of these numbers were taken and the
corresponding observations were removed from the data. The model was
run on the data with these observations missing. We then used the model
results to predict values for our missing observations.

\begin{figure}[b]

\includegraphics{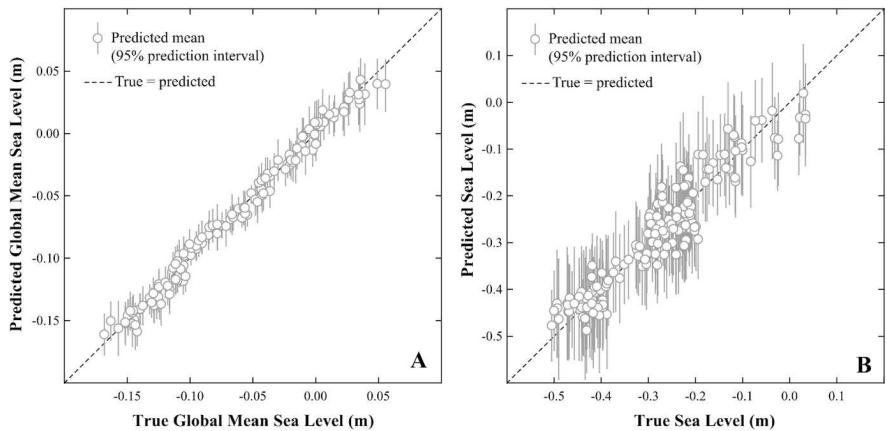}

\caption{True vs Predicted. \textup{(A)}  results for Church and
White (\citeyear{CandW2011}) global tide-gauge compilation.
\textup{(B)} shows the results for the North Carolina proxy reconstruction.}
\label{fig5}
\end{figure}

By definition, the observed data is an integral of the rate process.
Therefore, by integrating the rate process that we obtained from
running the model, at the points where we had missing data, we obtained
predictions for our missing data. We took sample paths from the
posterior estimates of the rate process (we used 500 samples) and
integrated these sample rate curves at the places where we removed the
observations. This provided us with 500 samples of posterior
predictions for each of the missing data points. From these samples we
determined the mean and standard deviations for each prediction and
used this information to approximate a mean prediction and a 95\%
prediction interval. Note, there is some variation in the prediction
intervals from sample to sample due the credible intervals for the rate
process varying slightly depending on which samples were removed.

\begin{figure}

\includegraphics{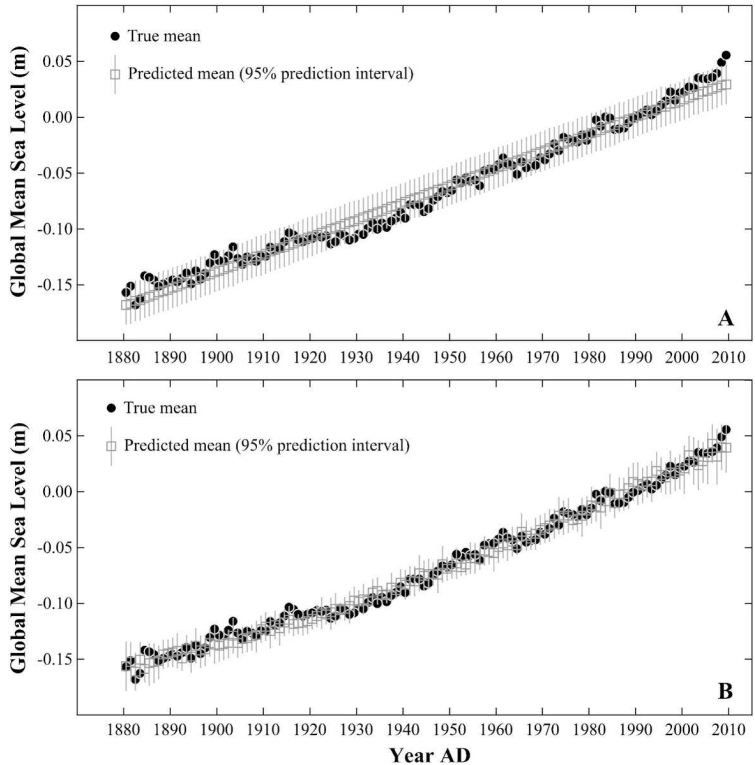}

\caption{Cross-validation for Church and White. Panel \textup{(A)} shows
predicted means for each sea-level observation and their 95\%
prediction intervals, overlaid on the true data, using a LSR modeling
approach. Panel \textup{(B)} shows the same results using an S-IGP approach.}
\label{fig6}
\end{figure}

\begin{figure}[t]

\includegraphics{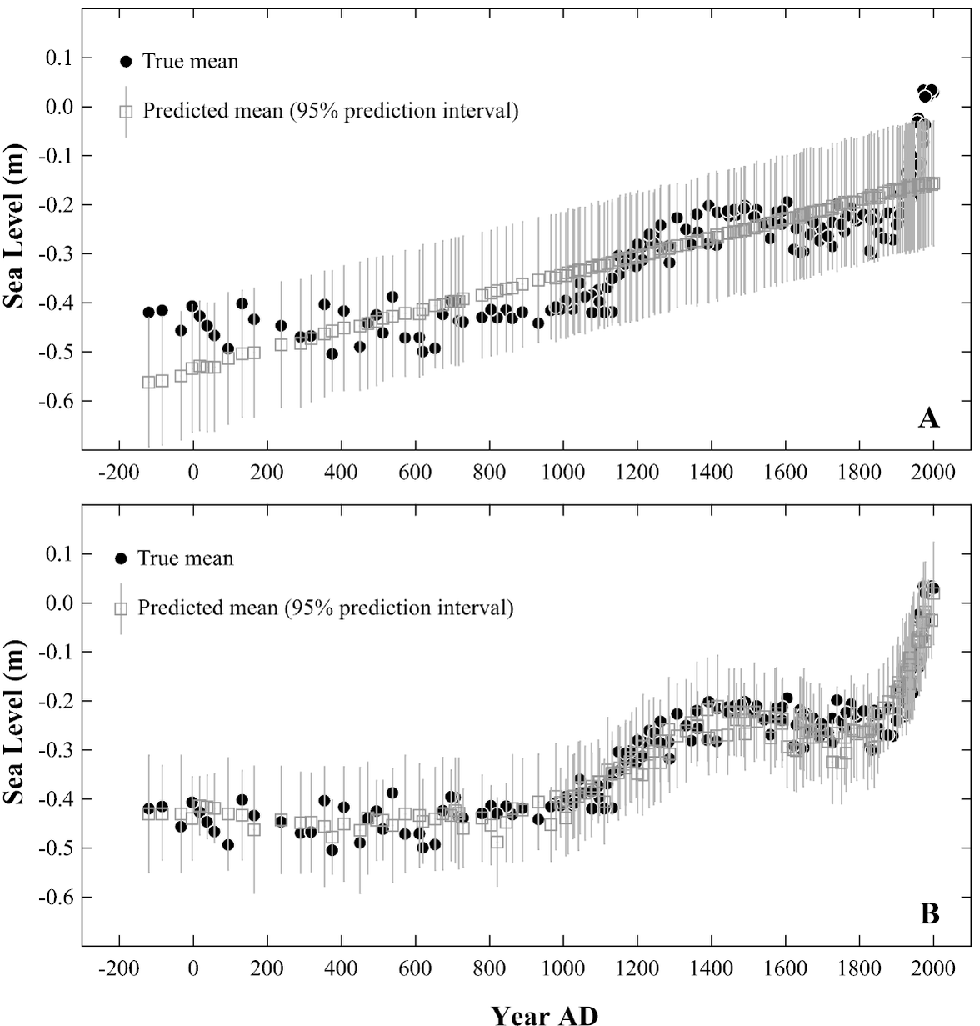}

\caption{Cross-validation for North Carolina. Panel \textup{(A)} shows predicted
means for each sea-level observation and their 95\% prediction
intervals, overlaid on the true data, using a LSR modeling approach.
Panel \textup{(B)} shows the same results using an
EIV-IGP approach.}\label{fig7}
\end{figure}

The cross-validation was carried out for both case study data sets. In
Figure~\ref{fig5} we plot the true sea-level observations versus the posterior
estimates of the mean predictions and their 95\% prediction intervals.
We performed the cross-validation for our IGP models in comparison to a
simplistic least squares regression (LSR) model for both case studies.
The results are displayed in Figures~\ref{fig6} and~\ref{fig7} for the global
tide-gauge data set of Church and White (\citeyear{CandW2011}) and the North Carolina
proxy measurements of Kemp et al. (\citeyear{Kemp2011}), respectively.

To asses the prediction intervals, we used the interval score in
equation \eqref{eqA3} proposed by \citet{predscore}:
%
\begin{equation}\label{eqA3}
S_{\alpha}^{\mathrm{int}}(l,u;x) =(u-l)+\frac{2}{\alpha}(l-x)1\{x<l\}+
\frac
{2}{\alpha}(x-u)1\{x>u\},
\end{equation}
where $l$ and $u$ are the upper and lower bounds of the
prediction interval and $x$ is the true value we are trying to predict.
The resulting score is negatively oriented and we wanted to minimize
the result. The score rewards for narrow prediction intervals and
penalizes if the observation misses the interval. Table~\ref{tab2} shows the
empirical coverage (\% of time the true observation fell within the
prediction interval), the average width of all the prediction intervals
and the average interval score for the prediction intervals.

\begin{table}[b]
\tabcolsep=0pt
\caption{Scoring results for the 95\% prediction intervals estimated
for each observation in the 10-fold cross-validation}\label{tab2}
%
\begin{tabular*}{\textwidth}{@{\extracolsep{\fill}}lcccc@{}}
\hline
\textbf{Data}&\textbf{Model}&\textbf{Empirical coverage}&\textbf{Average interval width}&\textbf{Average interval
score}\\
\hline
C$\&$W & LSR &97.69\%& 0.035&0.039\\
C$\&$W & S-IGP&95.34\% & 0.027& 0.032\\
N. Carolina & LSR &93.53\%&0.256 &0.340\\
N. Carolina &EIV-IGP&95.27\% & 0.182&0.198\\
\hline
\end{tabular*}
\end{table}
\end{appendix}

\newpage

\section*{Acknowledgments}
We are grateful to the Editor Tilmann Gneiting, the Associate Editor
and the two anonymous reviewers for their comments that greatly
improved the early version of the paper. We would also like to thank
Professor John Haslett for his comments and suggestions
and Professor Adrian Raftery for his help with the interval scoring rules.

\begin{supplement}[id=suppA]
\stitle{Data and code}
\slink[doi]{10.1214/15-AOAS824SUPP} 
\sdatatype{.zip}
\sfilename{aoas824\_supp.zip}
\sdescription{We provide the tide-gauge and proxy reconstructed data
for both case studies. We also supply the R code and JAGS code needed
to run the S-IGP and EIV-IGP models described.}
\end{supplement}

%
%

%



\printaddresses
\end{document}